\documentclass[prb,twocolumn,superscriptaddress,showpacs,amsmath,amssymb]{revtex4}

\usepackage{graphicx}
\usepackage{latexsym}
\usepackage{amsmath}
\usepackage{amssymb}
\usepackage{amsfonts}
\usepackage{color}
\usepackage{bm}
\usepackage{verbatim}

\begin{document}
\title{On the possibility of a long range proximity effect in a ferromagnetic nanoparticle.}

\author{M.~A.~Silaev}
\affiliation{Institute for Physics of Microstructures, Russian
Academy of Sciences, 603950 Nizhny Novgorod, GSP-105, Russia}

\date{\today}

\begin{abstract}
We study the proximity effect in a ferromagnetic nanoparticle
having a vortex magnetization pattern. We show that for
axisymmetric system consisting of a circular particle and a
magnetic vortex situated at the center of it no long range
superconducting correlations are induced. It means that induced
superconductivity is localized in the small area near the
superconducting electrode. However, in the real systems axial
symmetry can be broken by either a shift of the magnetic vortex
from the origin or geometrical anisotropy of the ferromagnetic
particle. In this case a long range proximity effect is possible.

\end{abstract}

\pacs{} \maketitle
\section{Introduction}

Proximity effect in hybrid ferromagnetic/superconducting (FS)
structures reveals a reach physics originating from the interplay
between magnetic and superconducting types of ordering (see
Ref.\onlinecite{BuzdinRMP} for review). There are two essential
features of the proximity effect in FS structures which make it
different from that in superconductor/ normal metal (SN)
structures. In SN structures the penetration length of a
condensate wave function into the normal metal is determined by
the normal metal coherence length $\xi_N=\sqrt{D/2\pi T}$, where
$D$ and $T$ are the diffusion coefficient and temperature. In
contrast a ferromagnetic coherence length which is also a depth of
the condensate penetration into a ferromagnet in FS system is much
shorter $\xi_F=\sqrt{D/h}$ provided the exchange energy $h$ is
rather large $h\gg T$ which is usually fulfilled.

Secondly, the
 penetration of Cooper pair wave
function into the ferromagnetic region (F) is characterized by the
damped oscillatory behaviour of a correlation function $f\sim
\exp(-x/\xi_F)\sin(x/\xi_F)$ which is a result of exchange
splitting between energy bands of conduction electrons with
different spin projections. In fact the origin of oscillations is
the same as for the Fulde- Ferrel- Larkin -Ovchinnikov
state\cite{FFLO}.
 This results in many new effects,
such us spatial oscillations of the density of
states\cite{DOSoscill}, a nonmonotonic\cite{TcNonMonotonic} or
reentrant\cite{TcReentrant} behaviour of the critical temperature
as a function of a ferromagnetic layer thickness in layered FS
structures. Also it is responsible for the formation of Josephson
$\pi$ junctions\cite{Josephson} and spin valves\cite{SpinValves}.

Despite the short coherence length in the ferromagnetic region
there is a possibility of a long range proximity effect in FS
structures with inhomogeneous magnetic structure. In experiments
on FS systems with strong ferromagnets an anomalously large
increase of the conductance below the superconducting critical
temperature $T_c$ was observed \cite{Experiment1, Experiment2,
Experiment3}. Also recently the Andreev interferometer geometry
was used to measure the phase sensitive conductance modulation in
the FS system with helical magnetic
structure\cite{PetrashovAndrInterf}.

The first theoretical analysis of a long range proximity effect in
FS structure with inhomogeneous magnetization was done  for a
Bloch-type domain wall at the FS interface \cite{BVEprl}. It was
shown that a superconducting correlation function contains
components which survive at the distances of order of the normal
metal correlation length from the superconducting boundary. These
long range superconducting components have non-trivial structure
in spin space. Conversely to the ordinary Cooper pairs which have
a singlet spin structure they have a triplet spin structure which
corresponds to correlations between electrons with the same spin
projections. Therefore the long range superconducting components
in FS systems are usually called the long-range triplet components
(LRTC).
 The LRTC can be generated in systems with
Bloch\cite{BVEprl} and Neel\cite{Neel1, Neel2} domain walls or
helical magnetization pattern\cite{Helical}. The long range
proiximity effect was shown to exist in multilayered FS structures
with noncollinear magnetization in different ferromagnetic
layers\cite{Multilayered1,Multilayered2,Multilayered3,Multilayered4}.
Large attention has been paid to the investigation of long range
Josephson effect due to LRTC in FS systems [see
Ref.\onlinecite{Efetov} for a review]. Recently in
Ref.\onlinecite{Multilayered4} a multilayered SFIFIFS structure
has been shown to demonstrate a controllable crossover between
long range triplet and short range singlet Josephson effects with
the rotation of the magnetic moment of any of the F layers.

The present paper is devoted to another possibility of
controllable switching between long and short range proximity
effects by employing the peculiar properties of ferromagnetic
nanoparticles. In some sense the magnetization of a nanoparticle
is more simple than the domain structure of macroscopic
ferromagnets, therefore, theoretical findings could be proved by
experiments with nanoparticles. It is now well-understood that a
magnetization distribution in a single particle is determined by
the competition between the magnetostatic and exchange energies.
If a particle is small, it is uniformly magnetized and if its size
is large enough a non-uniform (vortex) magnetization is more
energy preferable (see, for example, Refs. \onlinecite{Novosad4,
From_Frayerman_JMMM1, From_Frayerman_JMMM2, Novosad5, Novosad1,
Novosad2,Novosad3,CowburnBuckle}). Besides the geometrical form
and size, the state of the particle depends on many other factors.
For example, if the ferromagnetic particle is initially in the
vortex state then by applying a homogeneous in-plane magnetic
field one can shift the center of a magnetic vortex towards the
particle edges\cite {Guslienko}. If the magnetic field is large
enough the magnetic vortex annihilates, i.e. the particle becomes
homogeneously magnetized. Conversely, applying magnetic field to
the homogeneously magnetized particle in the direction opposite to
its magnetic moment one can force a nucleation of magnetic vortex.
Experimentally the shifting of magnetic vortex is observed as a
linear growth of the particle magnetic moment which saturates at
the field of vortex annihilation. A transition from homogeneous to
vortex state leads to a large jump of the magnetic moment so the
magnetization curve of a ferromagnetic nanoparticle is in general
highly hysteretic\cite{MagnetizationCurve}.

In practice superconducting correlations in a ferromagnetic
nanoparticle can be induced in planar geometry by lateral
superconducting junctions connected to the particle. Obviously if
the particle is homogenously magnetized then no long range
correlations are induced and the proximity effect is short range.
If the particle is in vortex state the situation is not so obvious
and the special investigation is needed. Throughout this paper we
will consider only the vortex state of the ferromagnetic particle.
We will show that for a circular particle there is no long range
superconducting correlations if the magnetic vortex is situated at
the center. However if the magnetic vortex is shifted from the
center by an external field there appear long range correlations.
Moreover an axial anisotropy of geometric form of the particle
also leads to a long range proximity effect.

The structure of this paper is following. In the next section we
describe our model, present the basic equations and give a
qualitative explanation of the long range proximity effect in a
ferromagnetic nanoparticle with vortex magnetization. In Section
III we present our main results which are discussed in Section IV.
Finally the conclusions are given in Section V.

 \section{Model and basic equations}
 \begin{figure}[h!]
\centerline{\includegraphics[width=1.0\linewidth]{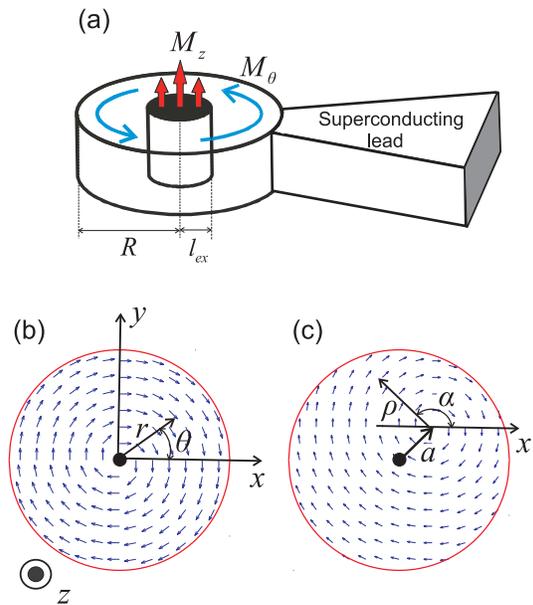}}
  \caption{\label{model}Sketch of the system considered:
  (a) Ferromagnetic nanoparticle with vortex magnetization
  and attached superconducting electrode.
 (b) Magnetic vortex at the center of the circular particle and the
  polar coordinate system. (c) Shifted magnetic vortex and the polar coordinate system with the origin at the
  vortex center. }
\end{figure}

We consider a system shown schematically in Fig.(\ref{model}). It
consists of a ferromagnetic nanoparticle and a lateral
superconducting lead. The particle magnetization is assumed to
form a magnetic vortex state. The structure of magnetic vortex is
shown in Fig. (\ref{model}b). It can be described by the rigid
vortex model proposed by Usov and Peschany\cite{UsovPeschany} and
by Guslienko\cite{Guslienko}. Within this model magnetization has
a $z$ component only inside the core region which size is
determined by a ferromagnetic exchange length $l_{ex}$. Outside
this region magnetization lies within $xy$ plane. Typically the
exchange length is quite small $l_{ex}\sim 10 nm$ compared to the
sizes of ferromagnetic nanoparticles $R\sim 100 nm$ therefore we
will neglect the vortex core region throughout this paper. Thus if
the center of magnetic vortex is situated at the point ${\bf
r}={\bf a}$ the magnetization distribution can be written in the
following form:
 \begin{equation}\label{MagnetizationShV1}
  {\bf M}=q_m M_0[{\bf n_\rho},{\bf z_0}],
 \end{equation}
 where ${\bf z_0}$ is a unit vector along $z$ axis,
 ${\bf n_\rho}=({\bf r}-{\bf a})/|{\bf r}-{\bf a}|$
  and ${\bf a}$ is a shifting vector of magnetic vortex center with respect to the origin.
  In polar coordinate system $(\rho,\alpha)$ with the origin at the center of
  magnetic vortex (see Fig.\ref{model}c) the magnetization
  distribution (\ref{MagnetizationShV1}) takes a simple form:
 \begin{equation}\label{MagnetizationShV}
  {\bf M}=q_m M_0(\sin\alpha, - \cos\alpha).
 \end{equation}
   The equation
 (\ref{MagnetizationShV})
 describes the magnetization vector curling around the center ${\bf r}={\bf a}$
 in a clockwise (counterclockwise) direction for $q_m=+1(-1)$.
 Further we will assume a clockwise direction of magnetization
 rotation (see Fig.\ref{model}b). Note that in case of a circular
 ferromagnetic particle the shift of a magnetic vortex from the
 center can be directly related to the external magnetic field ${\bf H_{ext}}$
 as follows:
 \begin{equation}\label{a(H)}
  {\bf a}=\chi_p\frac{{\bf z_0}\times {\bf H}}{M_0},
 \end{equation}
  where $\chi_p$ is the ferromagnetic nanoparticle linear magnetic
 susceptibility \cite{Guslienko}.
 The corresponding distribution of the effective exchange field acting on free
 electrons can be taken as ${\bf h}=h_0 {\bf M}/M_0$, where $h_0$
 is determined by the value of the exchange integral (see e.g. Ref.\onlinecite{BuzdinRMP}).

Our goal is to find a condensate Green function in the
ferromagnetic particle induced by an attached superconducting lead
due to a proximity effect (see Fig.\ref{model}a). We consider the
"dirty limit" assuming that the mean free path of electrons is
much shorter than all coherence lengths: $l\ll \xi_s,\xi_N,\xi_F$.
The most restrictive condition is $l\ll\xi_F$ since the
ferromagnetic coherence length is much shorter than coherence
lengths in superconductor $\xi_s$ and normal metal $\xi_N$. It
imposes certain limitation on the magnitude of exchange
interaction which means that the ferromagnetic should not be very
"strong".

To analyze a proximity effect in ferromagnetic particle we will
use Usadel equations for quasiclassical Green functions. Following
the scheme presented in detail in review \cite{Efetov} we
introduce a matrix Green function\cite{BookKopnin}
 $$
 \check{g}=\begin{pmatrix}
   \hat{g} & \hat{f} \\
   \hat{f}^+ & \bar{\hat{g}} \
 \end{pmatrix}.$$
  Here $\hat {g}$ is normal and $\hat{f}$ is anomalous
 Green functions which are matrices in spin space. A space where the matrix
 $\check{g}$ is defined is a Gor'kov-Nambu space. We will denote Pauli matrices
in Gor'kov-Nambu space as
 $\hat{\tau}_i$ and in spin space as $\hat{\sigma}_i$ ($i=1,2,3$).
 Unit matrices are $\hat{\tau}_0$ and $\hat{\sigma}_0$ correspondingly.
 Following Ref.\onlinecite{Efetov} the spinor basis for Green functions is taken
 in the following form:
  $$
  \hat{g}=\begin{pmatrix}
   \uparrow\uparrow & \uparrow\downarrow \\
   \downarrow\uparrow & \downarrow\downarrow \
 \end{pmatrix},
  $$
 $$
  \hat{f}=\begin{pmatrix}
   \uparrow\downarrow & \uparrow\uparrow \\
   \downarrow\downarrow & \downarrow\uparrow \
 \end{pmatrix},
  $$
  where $\uparrow$ and $\downarrow$ denote the spinors corresponding to spin projections
   $s_z=\pm  1/2$.

  It is convenient to use a transformation of Green function $\check{g}$ suggested by Ivanov and
  Fominov \cite{Fominov} $\check{g}=\check{V}\check{g}_{new}\check{V}^+$, where
 \begin{equation}\label{IvanovFominov}
 \check{V}=\exp\left(-i\frac{\pi}{4}(\hat{\tau}_3-\hat{\tau}_0)\hat{\sigma}_3\right).
 \end{equation}
 After this transformation is done the Usadel equation for the matrix
 Green function $\check{g}$ takes the following form:
   \begin{equation}\label{usadel}
  D\nabla(\check{g}\nabla\check{g})-\omega\left[\hat\tau_3,\check{g}\right]-
  i\left[\hat{\tau}_3({\bf h}\cdot{\bf \hat{\sigma}}),\check{g}\right]-[\check{\Delta},\check{g}]=0,
 \end{equation}
 where $[..]$ is a commutator, $D$ is a diffusion coefficient, $\omega$
 is Matsubara frequency and ${\bf h}= (h_x,h_y,h_z)$ is an effective exchange field.
 The gap function is given by
  $$
  \check{\Delta}=(\hat{\tau}_1Im\Delta-\hat{\tau}_2Re\Delta)\hat{\sigma}_0,
  $$
 If there are no superconducting correlations in the normal metal region
 then the Green function (in Matsubara representation) is
 given by
 \begin{equation}\label{NormalMetal}
  \check{g}(\omega)=sgn(\omega) \hat\tau_3\hat\sigma_0.
 \end{equation}
 The Eq. (\ref{usadel}) can be linearized assuming that
 \begin{equation}\label{lin}
  \check{g}=sgn(\omega)\hat\tau_3\hat\sigma_0+\check{F},
\end{equation}
 where second term is small $|\check{F}|\ll 1$.
    Then we obtain a linearized equation for $\check{F}$:
  \begin{equation}\label{LinUS}
 D\nabla^2\check{F}-2|\omega|\check{F}-
 i sgn(\omega) \{\hat{\tau}_0({\bf h}\cdot{\bf \hat \sigma}),\check{F}\}=0,
 \end{equation}
 where $\{..\}$ is anticommutator.
 The linearized boundary condition for the function $\check{F}$ at the S/F
interface is:
\begin{equation}\label{bcF}
  {\bf n}\cdot\nabla\check{F}=\check{F}_S/\gamma,
\end{equation}
 where $\gamma=R_b\sigma,$ while $R_b$ is the interface resistance
 per unit area and $\sigma$ is the conductivity of the
 ferromagnet, ${\bf n}$ is a unit vector normal to boundary.
 The anomalous function in bulk
 superconductor is:
 $$
 \check{F}_{S}
 =\left(\hat{\tau}_1\sin\varphi-\hat{\tau}_2\cos\varphi\right)\hat{\sigma}_0F_{bcs}.
 $$
Here $F_{bcs}=\Delta_0/\sqrt{\Delta_0^2+\omega^2}$, where
$\varphi$ and $\Delta_0$ is the phase and module of the
superconducting order parameter.

Note that in Eq.(\ref{LinUS}) the
 components of $\check{F}$ proportional to $\hat{\tau}_1$ and
 $\hat{\tau}_2$ are not coupled to each other.  Thus in
 ferromagnetic region anomalous function has the following
 structure: $\check{F}=\left(\hat{\tau}_1\sin\varphi-\hat{\tau}_2\cos\varphi\right) \hat{f}$,
 where $\hat{f}$ is a matrix in spin space.  For the function $\hat{f}$
 (when matrices in Namby space omitted) we obtain the following
 equation in ferromagnetic region:
 \begin{equation}\label{LinUS2}
 D\nabla^2\hat{f}-2|\omega|\hat{f}-
i sgn(\omega) {\bf h}\cdot\{{\bf \hat \sigma},\hat{f}\}=0.
 \end{equation}

The solution of Eq.(\ref{LinUS2}) can be found as a superposition:
 \begin{equation}\label{Pauli}
  \hat{f}=a_0\hat{\sigma}_0+a_1\hat{\sigma}_1+a_2\hat{\sigma}_2+a_3\hat{\sigma}_3.
 \end{equation}
In this expansion the first term corresponds to the singlet
component and the last three terms correspond to the triplet
components of anomalous function with different directions of
Cooper pair spin. Note that after the transformation
(\ref{IvanovFominov}) the spin space basis for the anomalous
function $\hat f$ can be symbolically written as follows:
 $$
  \hat{f}=\begin{pmatrix}
   \uparrow\downarrow & -\uparrow\uparrow \\
   \downarrow\downarrow & -\downarrow\uparrow \
 \end{pmatrix}.
 $$
Therefore it can be seen
 $$
 \hat S_i \hat\sigma_i=0,
 $$
 where $\hat S_i$ is an operator of spin projection for a Cooper
 pair with respect to the $i$-th axis ($i=x,y,z$). If the
 vector ${\bf f_{tr}}=(a_1,a_2,a_3)$ is parallel to some real vector ${\bf
 q}$ in $3D$ space then the Cooper pair spin projection on the
 vector ${\bf q}$ is zero. This means that the Cooper pairs
 consist of electrons with the opposite spin projections, or in
 other worlds the spin lies in the plane perpendicular to vector
 ${\bf q}$. As we will see below the exchange field ${\bf h}$ collinear with the vector ${\bf q}$
 effectively decouples the electrons
 leading to the fast decay of Cooper pair wave function into the ferromagnetic region.
 Otherwise if the vector ${\bf q}$ (or more generally ${\bf f_{tr}}$)
  is not collinear to exchange field ${\bf h}$ the LRTC appear.

  The equations for coefficients $a_i$ are:
\begin{equation}\label{a00}
  D\nabla^2a_0-2|\omega|a_0-
 i sgn(\omega) {\bf h}\cdot {\bf f_{tr}}=0,
\end{equation}
\begin{equation}\label{a1}
  D\nabla^2a_1-2|\omega|a_1-
i sgn(\omega)  h_x a_0=0,
\end{equation}
\begin{equation}\label{a2}
  D\nabla^2a_2-2|\omega|a_2-
i sgn(\omega)  h_y a_0=0,
\end{equation}
\begin{equation}\label{a3}
 D\nabla^2a_3-2|\omega|a_3-
i sgn(\omega)  h_z a_0=0.
\end{equation}

Now let us discuss the general structure of solutions of Eqs.
(\ref{a00}, \ref{a1}, \ref{a2}, \ref{a3}).   If the magnetization
and thus exchange field are homogeneous than it is easy to see
that there are two types of solutions of Eqs. (\ref{a00},
\ref{a1}, \ref{a2}, \ref{a3}): (i) short range and (ii) long range
modes. Indeed if the vector ${\bf f_{tr}}=(a_1,a_2,a_3)$ is
  parallel to the vector ${\bf h}$ then we obtain two equations
  for the functions $a_0$ and $b=|{\bf f_{tr}}|$:
  \begin{equation}\label{a0ShR}
 D\nabla^2a_0-2|\omega|a_0-
i sgn(\omega)  h b=0,
\end{equation}
\begin{equation}\label{bShR}
 D\nabla^2b-2|\omega|b-
 i sgn(\omega)  h a_0=0
\end{equation}
which have solutions in the form:
 $(a_0,b)\sim \exp(\lambda {\bf n}\cdot{\bf r})$, where $\lambda=\pm(1\pm i)k_h/\sqrt{2}$  and
$k_h=1/\xi_F=\sqrt{h/D}$ and ${\bf n}$ is a unit vector with
arbitrary direction.  These modes are short range ones since
ferromagnetic exchange length $\xi_F$ is typically very short. One
can see that short range modes consist of the singlet part of the
anomalous function with the amplitude given by coefficient $a_0$.
Also the is a nonzero contribution from triplet parts. The Cooper
pair spin is directed perpendicular to the exchange field ${\bf
h}$. Therefore, such triplet superconducting correlations are
suppressed by the exchange field on the same length scale as the
singlet ones.

On the other hand if the vector ${\bf f_{tr}}$ is perpendicular to
${\bf h}$ then the Cooper pair spin can be oriented along
 ${\bf h}$. In this case the destructive influence of exchange
 field on Cooper pairs is reduced.
 Indeed, from Eqs.(\ref{a00}, \ref{a1}, \ref{a2}, \ref{a3})
  we obtain that $a_0=0$ and $b$ satisfies the following equation
  \begin{equation}\label{bLR}
  D\nabla^2b-2|\omega|b=0,
\end{equation}
which have a solution $b\sim \exp(\lambda {\bf n}\cdot{\bf r})$,
where $\lambda=\pm 1/\xi_N$ and $\xi_N=\sqrt{D/|\omega|}$. These
modes are long range ones because the coherence length in normal
metal $\xi_N$ can be rather large. Note that since $a_0=0$ these
modes contain no singlet component, i.e. they contain only LRTC.

In case of homogeneous magnetization long range modes can not be
excited because of the zero boundary conditions for the triplet
components:
\begin{equation}\label{bcShR}
{\bf n} \cdot\nabla a_i=0
\end{equation}
 for $i=(1,2,3)$. The sources at the FS boundary exist only for
a singlet component:
\begin{equation}\label{bc}
{\bf n} \cdot\nabla a_0=F_{bcs}/\gamma,
\end{equation}
  where ${\bf n}$ is a unit vector normal to the boundary.
However it is not so for the inhomogeneous magnetization
distribution. The well-known examples when LRTC can be excited are
Bloch domain wall in a thin ferromagnetic wire\cite{BVEprl} or
spiral magnetic structure which can be realized in some rare-earth
metals\cite{Helical}. Also recently the case of Neel domain walls
in planar proximity FS structure was investigated\cite{Neel1,
Neel2}.

 Now let us consider magnetic structure with large scale
 inhomogeneities. In zero order approximation
for short range modes we obtain the Eqs.(\ref{a0ShR},\ref{bShR})
for $a_0$ and $b=|{\bf f_{tr}}|$ again, although the direction of
vector ${\bf a}$ adaibatically depends on the coordinate: ${\bf
f_{tr}}=b{\bf h}/h$. The solution can be written in the following
form: $(a_0, b)=(A, B) F({\bf r})$, where $A$ and $B=A
(k_h/\lambda)^2$ are constant and $F({\bf r})= \exp({\bf
\lambda}\cdot{\bf r})$. The boundary conditions
(\ref{bcShR},\ref{bc}) can be written as follows:
 \begin{equation}\label{bca0NH}
 {\bf n}\cdot \nabla a_0=F_{bcs}/\gamma
  \end{equation}
  \begin{equation}\label{bcbNH}
 {\bf h}({\bf n}\cdot \nabla b)=-b\;({\bf n}\cdot \nabla) {\bf h},
  \end{equation}
   where ${\bf n}$ is a unit vector normal to the boundary. There are two short range modes which
   decay far from FS boundary in the ferromagnetic region, say with $\lambda_1=k_h(1+i)/\sqrt{2}$ and
    $\lambda_2=k_h(1-i)/\sqrt{2}$. Taking the superposition of these modes
 with arbitrary coefficients $A_1$ and $A_2$ we obtain from Eq.(\ref{bcbNH}):
 $$
 A_1(\lambda_1-S_1)+A_2(\lambda_2-S_2)=0
 $$
 $$
 A_1(\lambda_1-S_2)+A_2(\lambda_2-S_1)=0,
 $$
 where $S_1=({\bf n}\cdot \nabla) h_x/h$ and $S_2=({\bf n}\cdot \nabla)
 h_y/h$. This linear system of the homogeneous equations has a solution
 if and only if $S_1=S_2$, i.e.
 \begin{equation}\label{condition}
 ({\bf n}\cdot \nabla) h_x=({\bf n}\cdot \nabla)
 h_y.
 \end{equation}
 This condition is fulfilled only in some special cases.
 The most trivial of them is a homogeneous magnetization distribution.
 Another particular case when condition (\ref{condition}) is fulfilled is that of
 a circular ferromagnetic particle
 if the magnetic vortex is situated at the center of the particle.
 Indeed in this case $h_{x,y}$ depend only on $\theta$ and
 therefore $({\bf n}\cdot \nabla) h_{x,y}=(\partial/\partial r)
 h_{x,y}=0$. Otherwise if the magnetic vortex is shifted from the center or the
 particle shape is axially symmetric the condition (\ref{condition}) is not fulfilled. It means
 that taking into account the short range modes only one can not satisfy the
 boundary conditions and with necessity the long range modes are excited.

 The above qualitative description of the eigen mode structure
  is based on the assumption of adiabatically slow
  variation of magnetization and exchange field ${\bf h}$. On the other hand in
  boundary condition (\ref{bcbNH}) appears a derivative of ${\bf h}$
  which in fact is a source for long range modes. Below we will
  find the corrections to the above adiabatic structure of
  short range modes. We will show that even if these corrections
  are taken into account it is still not possible to satisfy
  boundary conditions (\ref{bcbNH}) considering only the short range
  modes.

\section{ Structure of Short- and long- range modes in magnetic vortex.}

 For further considerations it is convenient to introduce new
 functions $b_\pm=a_1\pm ia_2$. Taking the magnetization distribution
  in the form (\ref{MagnetizationShV}) we obtain:
\begin{equation}\label{a0General}
  \left(\nabla^2-k_\omega^2\right) a_0-
 i\frac{k_h^2}{2}sgn(\omega) \left[S^*({\bf r})b_+ +S({\bf r})b_- \right]=0,
\end{equation}
\begin{equation}\label{b+General}
  \left(\nabla^2-k_\omega^2\right)b_+-
 i sgn(\omega) k_h^2 S({\bf r})a_0  =0,
\end{equation}
\begin{equation}\label{b-General}
 \left(\nabla^2-k_\omega^2\right)b_--
 isgn(\omega)  k_h^2 S^*({\bf r})a_0 =0,
\end{equation}
where $k_\omega^2=2|\omega|/D$ and $k_h^2=h_0/D$. We have
 introduced the following function:
 $S({\bf r})=(h_x+ih_y)/h_0$, where $h_0=\sqrt{h_x^2+h_y^2}$.

\subsection{Short range modes.}

 Usually the ferromagnetic coherence length $\xi_F=1/k_h$
 is very short. Most importantly it is much smaller
 than the size of a particle $R$ and the characteristic scale of
 the magnetization distribution given by the function $S({\bf r})$.
 Therefore solutions of Eqs.(\ref{a0General},\ref{b+General},\ref{b-General})
 with effective wavelength $\xi_F$ can be described within
 quasiclassical approximation. Also we neglect terms proportional to
 $k_\omega^2$. Physically it is justified since usually the normal metal coherence length
 $\xi_N\sim 1/k_\omega$ is much larger than the ferromagnetic coherence length $\xi_F$.


 Then we obtain the solution of
 Eqs.(\ref{a0General},\ref{b+General},\ref{b-General})
 in the following form (see Appendix \ref{ShR} for details):

\begin{equation}\label{a0}
   a_0= F(\theta)\exp(\lambda\; {\bf n\cdot r})
 \end{equation}
 \begin{equation}\label{b+}
  b_+=i F(\theta)sgn(\omega)
 \exp(\lambda\; {\bf n\cdot r})\frac{k_h^2}{\lambda^2}\left(1
 -\frac{2}{\lambda}({\bf n\cdot \nabla})\right)S,
 \end{equation}
 \begin{equation}\label{b-}
 b_-=i F(\theta)sgn(\omega)
 \exp(\lambda\; {\bf n\cdot r})\frac{k_h^2}{\lambda^2}\left(1
 -\frac{2}{\lambda}({\bf n\cdot \nabla})\right)S^*,
 \end{equation}
 where $F(\theta)$ is an
 arbitrary $2\pi$ periodic function and
 $\lambda=\lambda_{1,2}=k_h(1\pm i)/\sqrt{2}$.

 \subsection{Long  range modes.}

Now we are going to consider slow modes of
Eqs.(\ref{a0General},\ref{b+General},\ref{b-General}). For this
purpose we choose the coordinate origin at the magnetic vortex
center $(\rho,\alpha)$ (see Fig.\ref{model}c). Then we have
$S({\bf r})=-ie^{i\alpha}$ and therefore
Eqs.(\ref{a0General},\ref{b+General},\ref{b-General}) allow
separation of variables: $a_{0}=a_{\rho 0}(\rho)e^{iM\alpha}$,
$b_+=b_{\rho +}(\rho)e^{i(M+1)\alpha}$, $b_-=b_{\rho
-}(\rho)e^{i(M-1)\alpha}$. Then we obtain:
 \begin{equation}\label{LRa0}
 \left[\frac{1}{\rho}\frac{\partial}{\partial \rho}\left(\rho\frac{\partial}{\partial
 \rho}\right)
 -\frac{M^2}{\rho^2}-k_\omega^2\right]a_{\rho 0}-
 sgn(\omega) \frac{k_h^2}{2}\left(b_{\rho -}-b_{\rho +}\right)=0,
 \end{equation}
\begin{equation}\label{LRb+}
\left[\frac{1}{\rho}\frac{\partial}{\partial
\rho}\left(\rho\frac{\partial}{\partial
 \rho}\right)
 -\frac{(M+1)^2}{\rho^2}-k_\omega^2\right]b_{\rho +}-
 sgn(\omega)  k_h^2a_{\rho 0}  =0,
\end{equation}
\begin{equation}\label{LRb-}
\left[\frac{1}{\rho}\frac{\partial}{\partial
\rho}\left(\rho\frac{\partial}{\partial
 \rho}\right)
 -\frac{(M-1) ^2}{\rho^2}-k_\omega^2\right]b_{\rho -}+
 sgn(\omega)  k_h^2a_{\rho 0} =0.
 \end{equation}

 The behaviour of solutions of Eqs.(\ref{LRa0},\ref{LRb+},\ref{LRb-}) depends
  on the ratio of the ferromagnetic particle size and
 normal metal coherence length $R/\xi_N$. Indeed if $R\gg \xi_N$
  these modes decay at the length $\xi_N$. This is not very interesting case both for the
experiment and for the theoretical study. Another limit which can
be investigated analytically is realized when $\xi_N\gg R$. It
means that the decay of the long range modes on the size of a
ferromagnetic particle is weak. This condition is the most
favorable for investigation of the long range proximity effect.
Therefore we neglect terms proportional to $k_\omega^2$ from
Eqs.(\ref{LRa0},\ref{LRb+},\ref{LRb-}).

 It is possible to find the long range modes as
expansion by the orders of small parameter
 $(Rk_h)^{-1}$. The
 details of the calculations are shown in Appendix \ref{LR}. We obtain
 the following solution: $a_{\rho 0}=0$,
 \begin{equation}\label{LRmodes}
 b_{\rho +}=b_{\rho -}=B\rho^{\sqrt{M^2+1}},
 \end{equation}
 where $B$ is an arbitrary
 coefficient.

\section{Results.}

We will find the distribution of anomalous Gor'kov function in a
ferromagnetic nanoparticle induced by a superconducting electrode
which is attached to the particle as it is shown in
Fig.(\ref{model}). The superconducting electrode attached at some
point to the ferromagnetic sample can be modeled by the
angle-dependent transparency of the FS interface
$\gamma=\gamma(\theta)$ in the boundary condition (\ref{bc}). For
simplicity we can consider a Gauss form of transparency:

\begin{equation}\label{Transparency}
  \gamma=\gamma_0\exp\left[-\frac{\left((\theta-\theta_0)\; mod\;
 2\pi\right)^2}{\delta\theta^2}\right] ,
\end{equation}
where $\delta\theta$ is determined by a junction width
$d=R(\delta\theta/2\pi)$.

Let us start with a general consideration. The boundary conditions
for the coefficients $a_0$, $b_+$, $b_-$ at the boundary of a
ferromagnetic particle read:
\begin{equation}\label{bca0}
  {\bf n\cdot \nabla} a_0=\frac{F_{bcs}}{\gamma (\theta)}
\end{equation}
\begin{equation}\label{bcb_pm}
  {\bf n\cdot \nabla} b_\pm=0.
\end{equation}

  To satisfy the
boundary condition for $a_0$ we
 take the superposition of solutions (\ref{a0},\ref{b+},\ref{b-})
 corresponding to $\lambda_1=k_h(1+i)/\sqrt{2}$ and $\lambda_1=k_h(1-i)/\sqrt{2}$
 with arbitrary functions $F_{1,2}(\theta)$. We will take into account only those
 solutions which decay far from the FS boundary. Using the
 expression (\ref{MagnetizationShV}) for the vortex  magnetization and taking into
 account that $S({\bf r})=-ie^{i\alpha}$ from Eqs.(\ref{bca0},\ref{bcb_pm})
   we obtain:
 \begin{equation}\label{R1}
  \sum_{j=1,2}\lambda_jF_j(\theta)=\frac{F_{bcs}}{\gamma},
 \end{equation}
 \begin{equation}\label{R2}
   e^{i\alpha}sgn(\omega)\sum_{j=1,2}F_j(\theta)\frac{k_h^2}{\lambda_j}\left[1-\frac{i}{\lambda_j}({\bf
 n}\cdot\nabla\alpha)\right]+{\bf n\cdot \nabla}b_{l+}=0
 \end{equation}
 \begin{equation}\label{R3}
 e^{-i\alpha}sgn(\omega)\sum_{j=1,2}F_j(\theta)\frac{k_h^2}{\lambda_j}\left[1+\frac{i}{\lambda_j}({\bf
 n}\cdot\nabla\alpha)\right]-{\bf n\cdot \nabla}b_{l-}=0,
 \end{equation}
 where $b_{l\pm} ({\bf r})$ are the contributions of the long range
 modes. The structure of the long range modes yields the following
 relation for the coefficients
 $e^{-i\alpha}b_{l+}=e^{i\alpha}b_{l-}$. Let us denote $sgn (\omega)b_{l0}=e^{-i\alpha}b_{l+}=e^{i\alpha}b_{l-}$.
  Then from Eqs.(\ref{R2},\ref{R3}) we obtain:
  \begin{equation}\label{R21}
 \frac{F_1}{\lambda_1}+\frac{F_2}{\lambda_2}=-ib_{l0}\frac{{\bf n}\cdot \nabla \alpha }{k_h^2},
 \end{equation}
   \begin{equation}\label{R31}
 ({\bf n}\cdot \nabla\alpha)
 \left(\frac{F_1}{\lambda_1^2}+\frac{F_2}{\lambda_2^2}\right)=i\frac{{\bf n}\cdot \nabla b_{l0}}{k_h^2}.
 \end{equation}
 One has $\lambda_{1,2}\sim k_h$, therefore the r.h.s. of Eq.(\ref{R21}) is
 small and with good accuracy
 $F_1/\lambda_1+F_2/\lambda_2=0$.
  The Eqs.(\ref{R1},\ref{R31}) then yield
 \begin{equation}\label{ShRamplitude}
  F_{1,2}=F_{bcs}/(2\gamma\lambda_{1,2})
 \end{equation}
 and
 \begin{equation}\label{bcB+}
  {\bf n\cdot \nabla}b_{l0}=
 \frac{i}{\sqrt{2}k_h}\frac{F_{bcs}}{\gamma}({\bf n\cdot \nabla}\alpha),
 \end{equation}

 We search the contribution from the long range modes as a superposition:
 \begin{equation}\label{LR0}
   b_{l0}= \sum C_m \rho^{\sqrt{m^2+1}}e^{im\alpha}
 \end{equation}
 where $\rho,\alpha$ are the polar coordinates relative to the
 center of a magnetic vortex.
We use numerical methods to calculate the coefficients in the sum
(\ref{LR0}). We assume the angle dependent transparency in
Eq.(\ref{bca0}) in the form (\ref{Transparency}) with
$\delta\theta=0.02$ and the value of the ferromagnetic coherence
length $\xi_F=0.02R$. Further we will consider two typical cases:
(i) magnetic vortex in a circular particle shifted from the center
of it and (ii) magnetic vortex at the center of a particle having
the elliptical shape.

 \subsection{Shifted magnetic vortex.}

 \begin{figure}[htb]
\centerline{\includegraphics[width=1.0\linewidth]{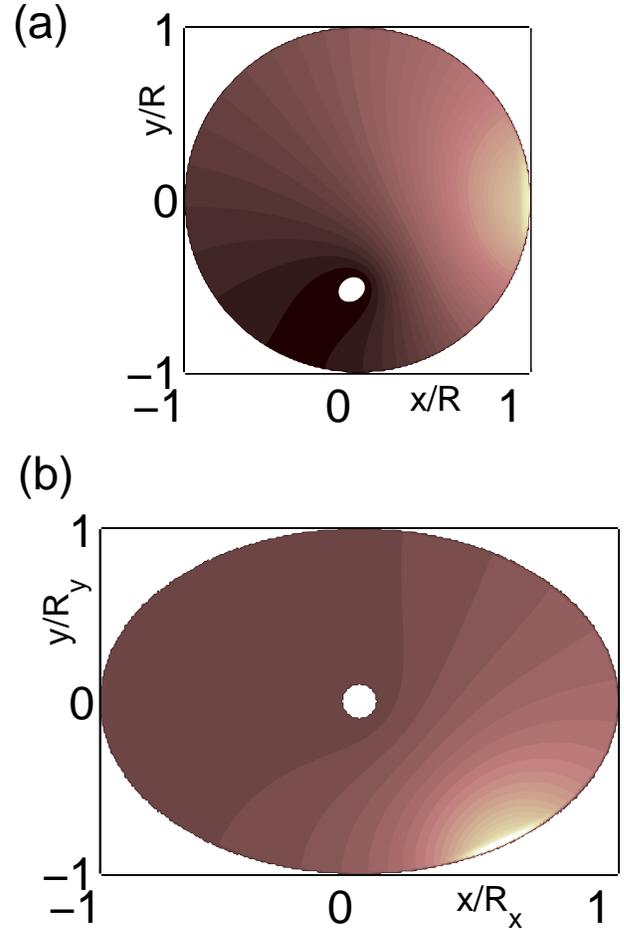}}
\caption{\label{ShiftedVortex}
  Amplitude of the triplet anomalous function induced
  due to the proximity effect. The position of magnetic vortex center is marked by the white circle.
   (a) Shifted magnetic vortex in circular
  particle, vortex shifting vector is ${\bf a}=(0,-R/2)$;
    (b) magnetic vortex at the center of elliptical
  particle with axes ratio $R_x/R_y=1.5$.}
  \end{figure}

 Let us assume for simplicity that the shifting vector is
 directed along $x$ axis: ${\bf a}=(a_x,0)$.
The vector normal to the boundary is directed along the disk
radius: ${\bf n}= {\bf r}/r$. Then the short range modes are given
by Eqs.(\ref{a0},\ref{b+},\ref{b-}) with
  $$
  S({\bf r})=
  \frac{i(a-re^{i\theta})}{\left(r^2-2ar\cos\theta+a^2\right)^{1/2}}
  $$
  and
  $$
  ({\bf n\cdot \nabla})S= \frac{dS}{dr}=\frac{a\sin\theta(a-r
  e^{i\theta})}{\left(r^2-2ar\cos\theta+a^2\right)^{3/2}}.
  $$
 The boundary condition for the long range modes (\ref{bcB+})
 takes the following form:
 \begin{equation}\label{bcB+ShV}
  \frac{db_{l0}}{dr}|_{r=R}=
 \frac{i}{\sqrt{2}k_h}\frac{F_{bcs}}{\gamma}Q(\theta),
 \end{equation}
 where
 $$
 Q(\theta)= \frac{a\sin\theta}{\left(R^2-2aR\cos\theta+a^2\right)}.
 $$
In general, the amplitudes of the short range modes given by
Eqs.(\ref{a0},\ref{ShRamplitude}) are determined by the
dimensionless factor $\xi_F/\gamma_0$. From the Eq.(\ref{LRmodes})
it is easy to see that $d b_{l0}/dr (r=R)\sim b_{l0}/R$. Thus,
when the vortex shifting is small ($a\ll R$) the amplitude of LRTC
is determined by the dimensionless factor $(\xi_F/\gamma_0)(a/R)$,
i.e. it is $(R/a)$ times smaller than the amplitudes of the short
range triplet components.

 In case when a junction with a superconducting lead is narrow
 $\delta\theta\ll 2\pi$, the amplitude of the LRTC is determined by the function
$|Q(\theta_0)|$. One can see that the maximum amplitude is
obtained when $\cos \theta_0= 2aR/(R^2+a^2)$. On the other hand
the long range proximity effect is absent if $\theta_0=0$ or
$\pi$. This is caused by the symmetry of the magnetization
distribution. In such case the magnetization is constant along the
direction of surface normal vector at the point where the
superconducting lead is attached. Therefore there appear no source
for LRTC at the FS boundary.

To demonstrate the enhancement of the LRTC in the ferromagnetic
particle with the shifted magnetic vortex we plot in
Fig.\ref{ShiftedVortex}a the distribution of the amplitude of the
triplet part of the anomalous function
 $|\hat f_{tr}|=\sqrt{|a_1|^2+|a_2|^2}$ [see expansion (\ref{Pauli})]. We
choose the position of a superconducting contact $\theta_0=0$ and
the magnetic vortex shifting vector ${\bf a}=(0,a_y)$.

\subsection{Magnetic vortex in elliptical particle}
Now let us consider the situation when the magnetic vortex is
situated at the center of a particle but the particle itself has
elliptical shape. The boundary of the elliptical particle is
determined by the equation $(x/R_x)^2+(y/R_y)^2=1$.
 It is convenient to write the vector normal to the boundary in the polar
 coordinate frame ${\bf n}=n_r {\bf r_0}+n_\theta {\bf \theta_0}$
 where
 $$
 n_r=\frac{R_y^2\cos^2\theta+R_x^2\sin^2\theta}{\sqrt{R_y^4\cos^2\theta+R_x^4\sin^2\theta}},
 $$
 $$
 n_\theta=\frac{(R_x^2-R_y^2)\sin(2\theta)}{2\sqrt{R_y^4\cos^2\theta+R_x^4\sin^2\theta}}.
 $$
 Then the short range modes are given
by Eqs.(\ref{a0},\ref{b+},\ref{b-}) with
 $S({\bf r})=-ie^{i\theta}$ and $({\bf n\cdot \nabla})S=  (n_\theta/r) (dS/d\theta)$,
    or
 $$
 ({\bf n\cdot \nabla})S=e^{i\theta}\sin(2\theta)\frac{R_x^2-R_y^2}{2R_xR_y}
 \sqrt{\frac{R_y^2\cos^2\theta+R_x^2\sin^2\theta}{R_y^4\cos^2\theta+R_x^4\sin^2\theta}}.
  $$
 The boundary condition for the long range modes (\ref{bcB+}) then
 takes the form (\ref{bcB+ShV}) with
 $$
 Q(\theta)=\sin(2\theta)\frac{R_x^2-R_y^2}{2R_xR_y}
 \sqrt{\frac{R_y^2\cos^2\theta+R_x^2\sin^2\theta}{R_y^4\cos^2\theta+R_x^4\sin^2\theta}}.
 $$
One can see that when the shape of the particle is nearly circular
$\delta R=\sqrt{|R_x^2-R_y^2|}\ll R_x,R_y$  the amplitude of LRTC
is determined by the dimensionless factor $(\xi_F/\gamma_0)(\delta
R/R_0)$, where $\delta R=\sqrt{|R_x^2-R_y^2|}$ is a measure of
axial anisotropy of the elliptical ferromagnetic nanoparticle and
$R_0=\sqrt{R_x^2+R_y^2}$.

Since the center of the magnetic vortex is assumed to coincide
with the particle center we search the long range modes in the
form of expansion (\ref{LR0}) with $\rho=r$ and $\alpha=\theta$.
Then we obtain:
 $$
 {\bf n}\cdot \nabla b_{l0}= \sum_m C_m
 r^{\sqrt{m^2+1}-1}e^{im\theta}
 \left(n_r\sqrt{m^2+1}+in_\theta m \right).
 $$

 Going along the same lines as in the previous section we find the
 coefficients $C_m$ numerically and obtain the distribution of the
amplitude of the triplet component of the anomalous function shown
in Fig.\ref{ShiftedVortex}b. We choose the position of the
superconducting contact $\theta_0=\pi/4$.

\section{Discussion}

 \begin{figure}[h!]
\centerline{\includegraphics[width=1.0\linewidth]{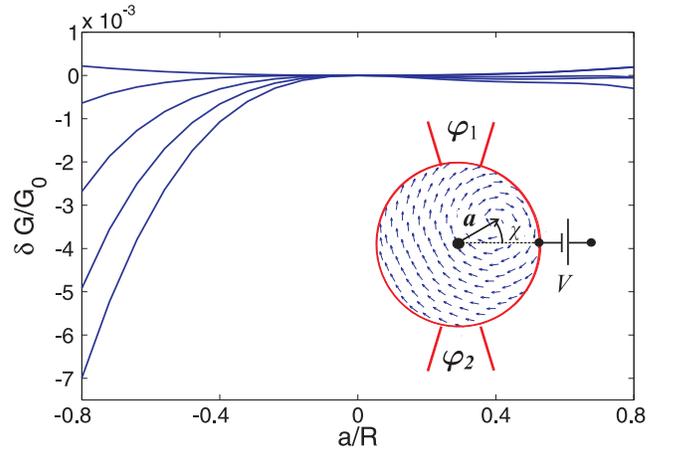}}
  \caption{\label{LDOS} A superconducting phase sensitive
  correction to the local conductance as a function of the magnetic
  vortex displacement with respect to the center of a circular
  ferromagnetic particle (see the insert). Different curves correspond to
  the angle $\chi$ values (from bottom to top): $\chi=0, \frac{2\pi}{10},
  \frac{3\pi}{10}, \frac{4\pi}{10},
  \frac{\pi}{2}$.}
    \end{figure}
Experimental observation of the proximity effect in FS structures
can be done for example using Andreev interferometer geometry to
measure the modulation of the conductivity of a ferromagnetic
sample as a function of the phase difference between the
superconducting leads\cite{PetrashovAndrInterf}. Therefore it is
interesting to investigate the influence of the long range
proximity effect on the transport properties of ferromagnetic
nanoparticles. Let us consider a system shown in Fig.\ref{LDOS}
(see the inset). We assume that there are two superconducting
leads with different phases of the superconducting order parameter
$\varphi_{1,2}$ attached at the different points to the circular
ferromagnetic nanoparticle. The normal lead measures the
conductance of the system. In case of a point junction with normal
lead one can use a general relation between a zero-bias tunneling
conductance $G$ and a local density of states (LDOS)  $\nu$ in the
ferromagnetic particle at the junction point:
\begin{equation}\label{CondLDOS}
G=G_n(\nu/\nu_n),
\end{equation}
 where $G_n$ and $\nu_n$ are the point junction conductance and LDOS in the normal state of
the ferromagnetic particle. The above expression for
 the local tunneling conductance is valid only if the voltage
 drops in the small vicinity of the junction point.
 This condition can be obtained assuming, for example, that the potential
 surface barrier is so high that all voltage drops just at the
 interface between the normal lead and the ferromagnetic particle.
 But in case of a point junction Eq.(\ref{CondLDOS}) can be used even for
  an ideal interface because the voltage
 drops at the distance determined by the junction size.
  Note that it is not so
 if, for example, a conductance of one-dimensional wire is considered\cite{BVEprl,VolkovSN}. We
 will assume that the junction size is much smaller than other characteristic
 lengths and employ the expression (\ref{CondLDOS}) for the tunneling conductance.

Having found the condensate function $\hat f$, we can calculate
the LDOS in the ferromagnetic region. The LDOS is given by the
general formula \cite{Efetov}:
 $$
 \nu=\frac{\nu_n}{4}Re Tr (\hat\tau_3\hat\sigma_0\check{g})
 $$
 where $\omega=-i\varepsilon+0$ and the trace is taken in both the
 Gor'kov-Namby and spin spaces.
 Using the normalization condition $\hat{g}^2+\hat f\hat f^+ =1$
 and the smallness of the condensate function, we obtain the correction to the
conductance of the point junction:
 $$
  \delta G/G_n= -\frac{1}{2}Re Tr (\hat f \hat f^+).
 $$
 The anomalous function $\check F$ has the following structure in Gor'kov-Nambu
 space:
  $$\check F= \sum _i\hat f_i \left( \sin\varphi_i\hat\tau_1-
   \cos\varphi_i\hat\tau_2\right),
 $$ where $\Delta_{1,2}=\Delta_0\exp(i\varphi_{1,2})$ are the gap functions in the superconducting
 leads. Therefore
 $$
 \hat f= i e^{-i\varphi_1}\hat f_1+i e^{-i\varphi_2}\hat f_2
 $$
 and
 $$
 \hat f^+= i e^{i\varphi_1}\hat f_1+i e^{i\varphi_2}\hat f_2.
 $$
  Thus we obtain:
\begin{equation}\label{ConductanceCorrection}
\delta G/G_n= \frac{1}{2}Re Tr \left(\hat f_1^2+\hat f_2^2+2\hat
f_1\hat f_2\cos\varphi\right),
\end{equation}
  where $\varphi=\varphi_1-\varphi_2$.
 Probably the most important for experiments is the conductance correction
 in Eq.(\ref{ConductanceCorrection}) which depends on the phase difference $\varphi$ due to the
 interference between the anomalous functions induced by different
 superconducting leads  $ \delta G/G_n=\cos\varphi Re Tr (\hat f_1\hat f_2) $.
 In Fig.\ref{LDOS} we show the
 dependence of the amplitude of conductance modulation $\delta G=G_n Re Tr (\hat f_1\hat f_2)$
 on the distance of
 the magnetic vortex center from the center of the ferromagnetic
 particle. Different curves in this plot correspond to the different directions of vortex shifting vector
 ${\bf a}$ (see the sketch of the system considered on the insert in Fig.\ref{LDOS}).
 We normalize the conductance to the
 following value $G_0=G_n(\gamma_0/\xi_F)$ which is entirely
 determined by the fixed parameters of the system.

 Analyzing
  Fig.\ref{LDOS} one can see that
  the strongest effect is achieved by shifting the
 vortex symmetrically with respect to the superconducting
 contacts ($\chi=0$). On the contrary,
  the effect of conductance modulation is very small in case of the vortex
 shifting along the line between two superconducting leads ($\chi=\pi/2$, top
 curve). As we have discussed above, in this case the LRTC are
 weak due to the symmetry of the magnetization distribution.
 A non-zero value of the conductance modulation in this case is caused
 only by the finite width of the superconducting junctions used in
 calculations. Furthermore, in Fig.\ref{LDOS} all the curves, except for
 the top one which corresponds to $\chi=\pi/2$, demonstrate strong asymmetry
 with respect to the sign of
 the vortex displacement $\delta G(a)\neq\delta G(-a)$.
 Such asymmetry is caused by the system
 geometry, since we consider a conductance of only one point junction.
 As one can see if the magnetic vortex shifts towards the normal contact
 (positive $a$ in Fig.\ref{LDOS}) the conductance modulation appears
 to be very small compared to the case when the magnetic vortex
 shifts in the opposite direction (negative $a$ in Fig.\ref{LDOS}).
 This effect can be understood
 if we recall that the long range modes are strongly suppressed near the vortex center
 [see Eq. (\ref{LRmodes})]. Thus even if the overall amplitude of
 LRTC is increased with $|a|$, the local value of anomalous function
 at the junction point is decreased if the magnetic vortex center
 shifts towards the junction point.

 The shift of the magnetic vortex is unambiguously determined by the magnetic field [see Eq.
 (\ref{a(H)})], therefore the asymmetry $\delta G(a)\neq\delta G(-a)$ will be revealed in the
 conductance dependence on the external magnetic field: $\delta G(H)\neq\delta
 G(-H)$. But in reality
 one always has two contacts and the total conductance correction
 is a sum of the contributions from each contact. Thus the
 resulting behaviour of the conductance should depend on
  the position of the points where superconducting and normal
contacts are connected to the ferromagnetic particle. In
particular, if the system geometry is symmetric with respect to
the spatial inversion the conductance correction will not depend
on the sign of vortex shifting $a$ as well as on the sign of the
magnetic field $\delta G(H)=\delta G(-H)$.

In Fig.\ref{LDOS} the modulation of conductance is shown not for
the entire range of the magnetic vortex displacements from the
particle center. The reason is a growing complexity of numerical
calculations because when the magnetic vortex center approaches
close to the particle boundary one has to take into account too
many angular harmonics in the expansion (\ref{LR0}). We expect
further monotonic growth of $|\delta G (a)|$ until $|a|<R$. If the
vortex displacement distance becomes larger than the particle
radius $|a|>R$, the vortex actually leaves the particle. Such
magnetization state often is referred as "buckle"
\cite{CowburnBuckle}. Further increase of $|a|$ describes in fact
a continuous transition to the homogeneously magnetized state.
Therefore, the conductance correction should eventually vanish as
$|a|\rightarrow 0$.

The overall magnitude of the conductance modulations is determined
by many factors. One of them is a vortex displacement, which can
be regulated by the external magnetic field. Other factors are
determined by the geometry of the system, e.g. width of
superconducting leads and the particle size $R$. Also there is a
dimensionless factor $\xi_F/\gamma_0$, which depends on the
material parameters: ferromagnetic coherence length $\xi_F$ and
$\gamma_0=R_F\sigma_{int}$, where $R_F$ is the resistance per unit
area of FS interface and $\sigma_{int}$ is the conductivity of
ferromagnetic\cite{Efetov}. This factor determines the amplitude
of the anomalous function within the ferromagnetic region and
should be small within our calculation scheme, because we consider
the linearized Usadel equation. For a particular configuration
shown on the inset in Fig.\ref{LDOS} we obtain the maximal
amplitude of conductance modulation $\delta G\sim 10^{-2}
(\xi_F/\gamma_0) G_n$, where $G_n$ is the unperturbed conductance
in the normal state of the particle. Taking for example
$\xi_F/\gamma_0\sim 10^{-2}$ we obtain that $\delta G\sim 10^{-4}
G_n$. To have a better effect in experiment one should try to
increase the ratio $\xi_F/\gamma_0$. For example this can be
obtained by using not very strong ferromagnetic material with
relatively large $\xi_F$ e.g., Cu-Ni alloys\cite{RyazanovCuNi},
characterized by rather large coherence lengths: $\xi_F\sim 10
nm$. However the magnetic vortex has been observed in rather
strong ferromagnets such as $Co$ or $Pe$ with $\xi_F\sim 1 nm$. On
the other hand, one can try to improve the properties of the
superconducting contacts, i.e. to use the contacts with low
interface resistance $R_F$.

\section{Conclusion}
To summarize we have investigated the proximity effect in the
ferromagnetic nanoparticle with nonhomogeneous vortex
magnetization distribution. We have derived a general solution
both for the short range components and the long range triplet
components of the anomalous function. Quite generally it is shown
that the long range proximity effect can be realized if the axial
symmetry of the magnetization distribution is broken either due to
the shifting of magnetic vortex with respect to the particle
center or due to the angular anisotropy of the particle shape,
which can be, for example, ellyptical in real experiments. Also we
have considered the superconducting phase-periodic oscillations of
the particle conductance in Andreev interferometer geometry, which
has been used recently to study the proximity effect in a conical
ferromagnet \cite{PetrashovAndrInterf}. We have shown that the
amplitude of conductance oscillations strongly depends on the
direction of external magnetic field which determines the shift of
magnetic vortex with respect to the particle center. For a
particular case of a circular ferromagnetic particle the
conductance oscillations are the largest when the vortex shifting
is symmetric with respect to the superconducting contacts
position. However, we suppose that the optimal direction of vortex
shifting for the observation of the long range proximity effect
should depend on the system geometry, such as particle shape and
position of the points where the superconducting and normal
contacts are connected to it.

\section{Acknowledgements}
I am grateful to A.A. Fraerman and A.S. Mel'nikov for drawing my
attention to this problem and for helpful discussions. This work
was supported, in part, by Russian Foundation for Basic Research,
by Program "Quantum Macrophysics" of RAS, and by Russian Science
Support and "Dynasty" Foundations.

\appendix
\section{Derivation of the short range modes}
\label{ShR}
 We search quasiclassical
  solutions of Eqs.(\ref{a0General},\ref{b+General},\ref{b-General}) in the following form:
\begin{equation}\label{QCansatz}
  (a_0,b_+,b_-)=\exp(\lambda\; {\bf n\cdot r})
 (a_{0q},b_{+q},b_{-q}),
\end{equation}
where ${\bf n}$ is a unit vector, $\lambda$ is large and functions
$a_{0q},b_{+q},b_{-q}$ are slow. Note that in principle the
direction of vector ${\bf n}$ is arbitrary and should be
determined from the boundary conditions. But we assume from the
beginning that the spatial scale of the anomalous function
variation along the boundary is much larger than $1/|\lambda|$.
Thus we can consider vector ${\bf n}$ as a normal to the boundary
of a ferromagnetic. Then at first we need to find $\lambda$.
Substituting functions in the form (\ref{QCansatz}) into
Eqs.(\ref{a0General},\ref{b+General},\ref{b-General}) we obtain:
$\lambda^4=-k_h^4$, i.e. $\lambda=(-1)^{1/4}k_h$ which corresponds
to the short range modes and $\lambda=0$ which we will discuss.
For quasiclassical envelopes we obtain the following equations:
 \begin{equation}\label{a0QC}
  2\lambda({\bf n \nabla})a_{0q}+\lambda^2a_{0q}-
 i\frac{k_h^2}{2}sgn(\omega) \left[S^*({\bf r})b_{+q} +S({\bf r})b_{-q} \right]=0,
\end{equation}
\begin{equation}\label{b+QC}
  2\lambda({\bf n \nabla})b_{+q}+\lambda^2b_{+q}-
 i sgn(\omega) k_h^2 S({\bf r})a_{0q}  =0,
\end{equation}
\begin{equation}\label{b-QC}
 2\lambda({\bf n \nabla})b_{-q}+\lambda^2b_{-q}-
 isgn(\omega)  k_h^2 S^*({\bf r})a_{0q} =0,
\end{equation}
Since all $\lambda$ have large real parts all the solutions decay
or grow very fast. We will take into account only those which
decay far from the boundary of the ferromagnetic particle. Then we
should leave $\lambda_1=k_h(1+i)/\sqrt{2}$ and
$\lambda_2=k_h(1-i)/\sqrt{2}$. Let us now find the solutions of
quasiclassical Eqs.(\ref{a0QC},\ref{b+QC},\ref{b-QC}). We will use
a perturbation method.

Let us at first assume that $a_{0q}=const$. Then to the zero
order:
 \begin{equation}\label{appB+0}
  b_{+q}=i a_{0q} sgn(\omega)  \frac{k_h^2}{\lambda^2}S({\bf r})
 \end{equation}
 \begin{equation}\label{appB-0}
  b_{-q}=i a_{0q} sgn(\omega)  \frac{k_h^2}{\lambda^2}S^*({\bf r}).
 \end{equation}
 Note that we also can assume $a_{0q}=G({\bf r})$, where $G({\bf
 r})$ is arbitrary but rather slow function. In this case two other coefficients $b_{+q}$
 and $b_{-q}$ are proportional to $G({\bf r})$. The condition
 $({\bf n \nabla}) G\ll |\lambda|$ guarantees that this will not change the
 structure of eigen modes.
 Substituting expressions (\ref{appB+0},\ref{appB-0}) into Eqs.(\ref{b+QC},\ref{b-QC}) we
 obtain the first order perturbations
 $$
  \tilde{b}_{+q}=-2ia_{0q} sgn(\omega)  \frac{k_h^2}{\lambda^3}({\bf n \nabla}) S({\bf
  r})
 $$
 $$
  \tilde{b}_{-q}=-2ia_{0q} sgn(\omega)  \frac{k_h^2}{\lambda^3}({\bf n \nabla}) S^*({\bf
  r}).
 $$

\section{Derivation of the long range modes}
\label{LR}

The long range modes can be found solving Eqs.
(\ref{LRa0},\ref{LRb+},\ref{LRb-}) with neglected terms
proportional to $k_\omega^2$:

 \begin{equation}\label{appLRa0}
 \left[\frac{1}{\rho}\frac{\partial}{\partial \rho}\left(\rho\frac{\partial}{\partial
 \rho}\right)
 -\frac{M^2}{\rho^2}\right]a_{\rho 0}-
 sgn(\omega) \frac{k_h^2}{2}\left(b_{\rho -}-b_{\rho +}\right)=0,
 \end{equation}
\begin{equation}\label{appLRb+}
\left[\frac{1}{\rho}\frac{\partial}{\partial
\rho}\left(\rho\frac{\partial}{\partial
 \rho}\right)
 -\frac{(M+1)^2}{\rho^2}\right]b_{\rho +}-
 sgn(\omega)  k_h^2a_{\rho 0}  =0,
\end{equation}
\begin{equation}\label{appLRb-}
\left[\frac{1}{\rho}\frac{\partial}{\partial
\rho}\left(\rho\frac{\partial}{\partial
 \rho}\right)
 -\frac{(M-1) ^2}{\rho^2}\right]b_{\rho -}+
 sgn(\omega)  k_h^2a_{\rho 0} =0.
 \end{equation}

 It is convenient to rearrange these equations
introducing new functions $b_s=b_{\rho +}+b_{\rho -}$ and
$b_d=b_{\rho +}-b_{\rho -}$:
  \begin{equation}\label{a0LRm}
 \left(\frac{\partial^2}{\partial \rho^2}+\frac{1}{\rho}\frac{\partial}{\partial
 \rho}-\frac{M^2}{\rho^2}\right)a_{\rho 0}+
 sgn(\omega) \frac{k_h^2}{2}b_d=0,
 \end{equation}
\begin{equation}\label{bsLR}
\left(\frac{\partial^2}{\partial
\rho^2}+\frac{1}{\rho}\frac{\partial}{\partial
 \rho}-\frac{M^2+1}{\rho^2}\right)b_s-\frac{2M}{\rho^2}b_d
   =0,
\end{equation}
\begin{equation}\label{bdLR}
 \left(\frac{\partial^2}{\partial \rho^2}+\frac{1}{\rho}\frac{\partial}{\partial
 \rho}-\frac{M^2+1}{\rho^2}\right)b_{d}-\frac{2M}{\rho^2}b_s-
 2 sgn(\omega)  k_h^2a_{\rho0} =0.
 \end{equation}
 We will find the solutions of these equations as expansion by the orders of the small parameter
 $(\rho k_h)^{-1}$ assuming that the distance from vortex center is much larger than the
  ferromagnetic coherence length $\rho\gg\xi_F$. It is easy to see that if
 $k_h\rightarrow\infty$,
  we obtain that $b_{d}=0$ and $a_{\rho 0}=0$ and
 \begin{equation}\label{bZO}
 \left(\frac{\partial^2}{\partial \rho^2}+\frac{1}{\rho}\frac{\partial}{\partial
 \rho}-\frac{M^2+1}{\rho^2}\right)b_s =0.
 \end{equation}
 The solution of this equation is $b_s=B_{s0}\rho^{\sqrt{M^2+1}}$.
Then from the Eq.(\ref{bdLR}) we get:
 $$
 a_{\rho 0}=-sgn(\omega)\frac{2M}{(k_h\rho)^2},
 $$

 $$
 b_s=-sgn(\omega)\frac{2M}{(k_h\rho)^2}\rho^{\sqrt{M^2+1}}B_{s0}.
 $$
The function $b_d$ is obtained from Eq.(\ref{a0LRm}):
 $$
 b_d=-\frac{4M(5-4\sqrt{M^2+1})}{(k_h\rho)^4}\rho^{\sqrt{M^2+1}}B_{s0}.
 $$
 Substituting it to the  Eqs.(\ref{bsLR}) we obtain the next correction for $b_s$ of the order
 $(k_h\rho)^{-4}$ which can be neglected.

\end{document}